\documentclass[11pt]{article}
\usepackage{moriond,epsfig}

\def\be{\begin{equation}}
\def\ee{\end{equation}}
\def\bea{\begin{eqnarray}}
\def\eea{\end{eqnarray}}
\def\smallfrac#1#2{\hbox{${{#1}\over {#2}}$}}

\begin{document}
\vspace*{1.9cm}
\title{The NNPDF2.2 Parton Set}

\author{ F. CERUTTI }

\address{Universitat de Barcelona,\\ Departament d'Estructura i Constituents de la Mat\`{e}ria,\\ Av. Diagonal 647,
Barcelona 08028, Espa\~{n}a.\\}
\author{ N.P. HARTLAND }
\address{School of Physics and Astronomy, \\University of Edinburgh, JCMB, KB, Mayfield Rd,\\ Edinburgh EH9 3JZ, Scotland.}

\maketitle\abstracts{We present a method developed by the NNPDF Collaboration that allows the inclusion of new experimental data into an existing set of parton distribution functions without the need for a complete refit. A Monte Carlo ensemble of PDFs may be updated by assigning each member of the ensemble a unique weight determined by Bayesian inference. The reweighted ensemble therefore represents the probability density of PDFs conditional on both the old and new data. This method is applied to the inclusion of W-lepton asymmetry data into the NNPDF2.1 fit producing a new PDF set, NNPDF2.2.
}

\section*{NNPDF Approach} 
The NNPDF methodology differs from other parton fitting approaches in two main aspects, the choice of parameterization and the treatment of uncertainties. Neural networks are used to parametrize the PDFs at the initial scale. These are chosen because they provide a robust, unbiased interpolation and permit a very flexible parameterization. An NNPDF fit has almost three hundred free parameters and is therefore free from the bias associated with the choice of an inflexible functional form. The flexibility of the neural network fit has been demonstrated in several ways. For example, the fit has been shown to be stable under changes in network architecture \cite{Ball:2011eq} and under the addition of new PDFs \cite{strange}.
\\\\
In order to obtain an accurate representation of PDF uncertainty, NNPDF fits are performed on pseudo data samples generated from the original experimental data by Monte Carlo sampling. Each sample, or 'replica' forms the basis for a separate fit, resulting in an ensemble of PDF replicas that faithfully represent the uncertainty in the experimental data, without the need for a tolerance criterion.
\\\\
These features of the NNPDF approach allow for an interesting exploitation of Bayesian inference to determine the impact of new datasets.

\section*{Reweighting Method}
We shall discuss how an existing probability distribution in the space of PDFs may be updated with information from new data. To include the new data, one can of course perform a fit with the new, enlarged dataset. However this is a time consuming task, particularly for observables where no fast code is available. It is therefore desirable to have a faster method of including new data in order to assess it's impact rapidly without the need for a full refit. NNPDF parton sets are supplied as an ensemble of $N$ parton distribution replicas $\mathcal{E}$, representing the probability density in PDFs $\mathcal{P}(f)$ based upon the data in the existing fit. We can therefore include new data by weighting each replica $f_k$ in the ensemble by an associated weight $w_k$. If the replica weights are computed correctly, then reweighting is completely equivalent to a refit.
\\\\
In order to illustrate the reweighting method, consider the computation of the expected value of a PDF-dependant observable $\mathcal{O}[f]$. We note that as the NNPDF Monte Carlo ensemble is a good representation of the probability density $\mathcal{P}(f)$, the expectation value $\left< \mathcal{O}[f] \right>$ can be calculated as a simple average,
$$
\langle\mathcal{O}\rangle=\int \mathcal{O}[f] \, \mathcal{P}(f)\,Df
=\smallfrac{1}{N}\,\sum_{k=1}^{N}\mathcal{O}[f_k]\, .
$$
New data can be included into the existing ensemble by assigning each replica a unique weight $w$. This weight assesses the agreement between the replica and the new data. The reweighted ensemble now forms a representation of the probability distribution of PDFs $\mathcal{P}_{\mathrm{new}}(f)$ conditional on both the existing and new data. The mean value of the observable $\mathcal{O}$ taking account of the new data is then given by the weighted average
$$
\langle\mathcal{O}\rangle_{new}=\int \mathcal{O}[f] \, \mathcal{P}_{new}(f)\,Df
=\smallfrac{1}{N}\,\sum_{k=1}^{N}w_k\mathcal{O}[f_k],\, 
$$
\\
where the weights are given in terms of the individual replica $\chi^2$ to new data by
$$
w_k = 
\frac{(\chi^{2}_k)^{(n-1)/2} 
e^{-\frac{1}{2}\chi^{2}_k}}
{\smallfrac{1}{N}\sum_{k=1}^{N}(\chi^{2}_k)^{(n-1)/2}
e^{-\frac{1}{2}\chi^{2}_k}}=\mathcal{N}_\chi\mathcal{P}(\chi^2|f_k)\, .
$$
\\
Note that after reweighting a given ensemble of $N$ PDF replicas we no longer have the same efficiency in describing the distribution of PDFs. The reweighting procedure will often assign replicas very small weights, therefore these replicas no longer contribute to the ensemble. The efficiency of the representation of the underlying distribution $\mathcal{P}_{new}(f)$ will therefore be less than it would be in a new fit. The loss of information due to reweighting can be quantified using the Shannon entropy to determine the effective number of replicas in the reweighted set:
$$
N_{\rm eff}\equiv \exp \{\smallfrac{1}{N}\sum_{k=1}^N w_k \ln (N/w_k)\}.
$$
 \begin{figure}[b!]
    \begin{center}
      \includegraphics[width=0.46\textwidth]{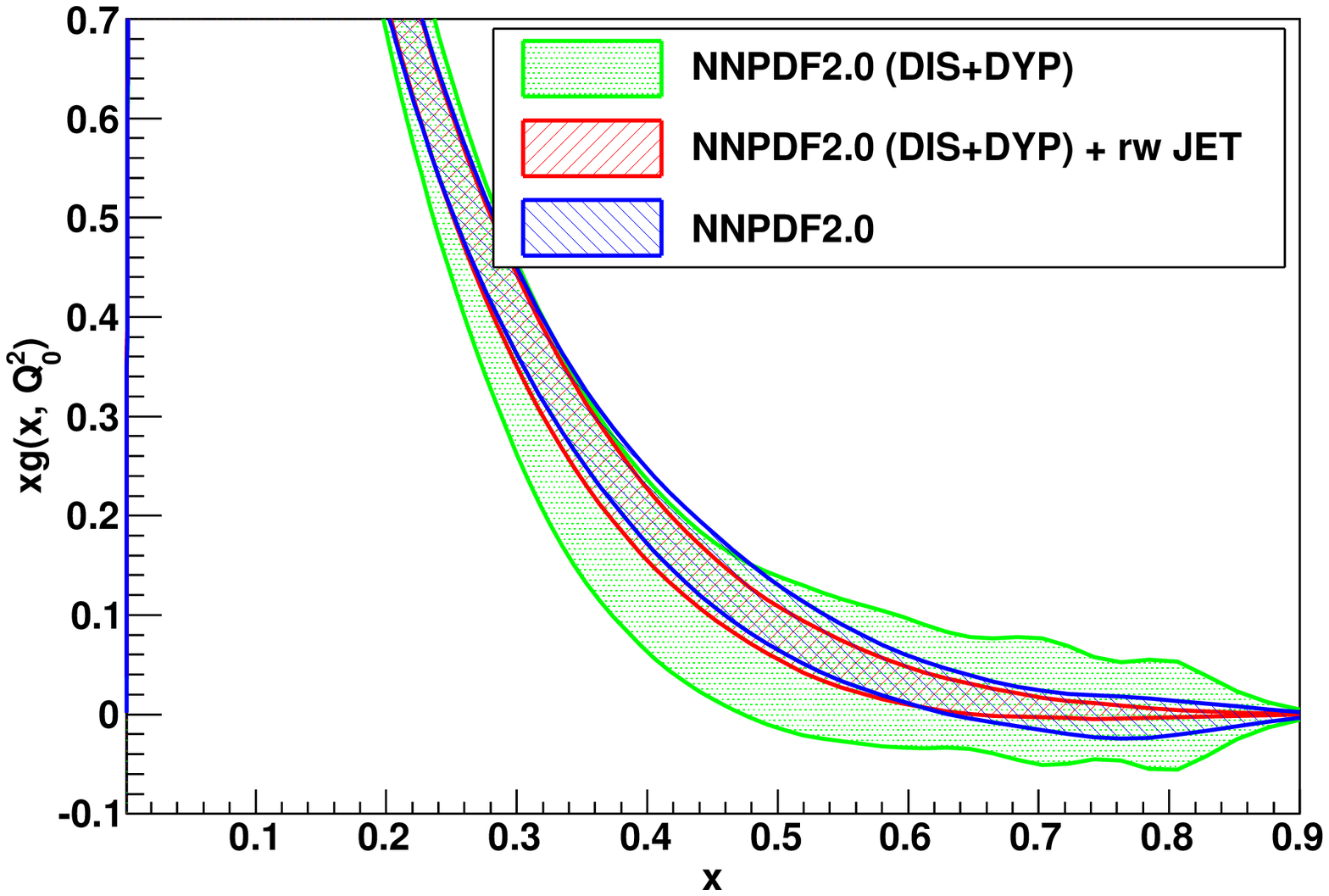}
      \includegraphics[width=0.46\textwidth]{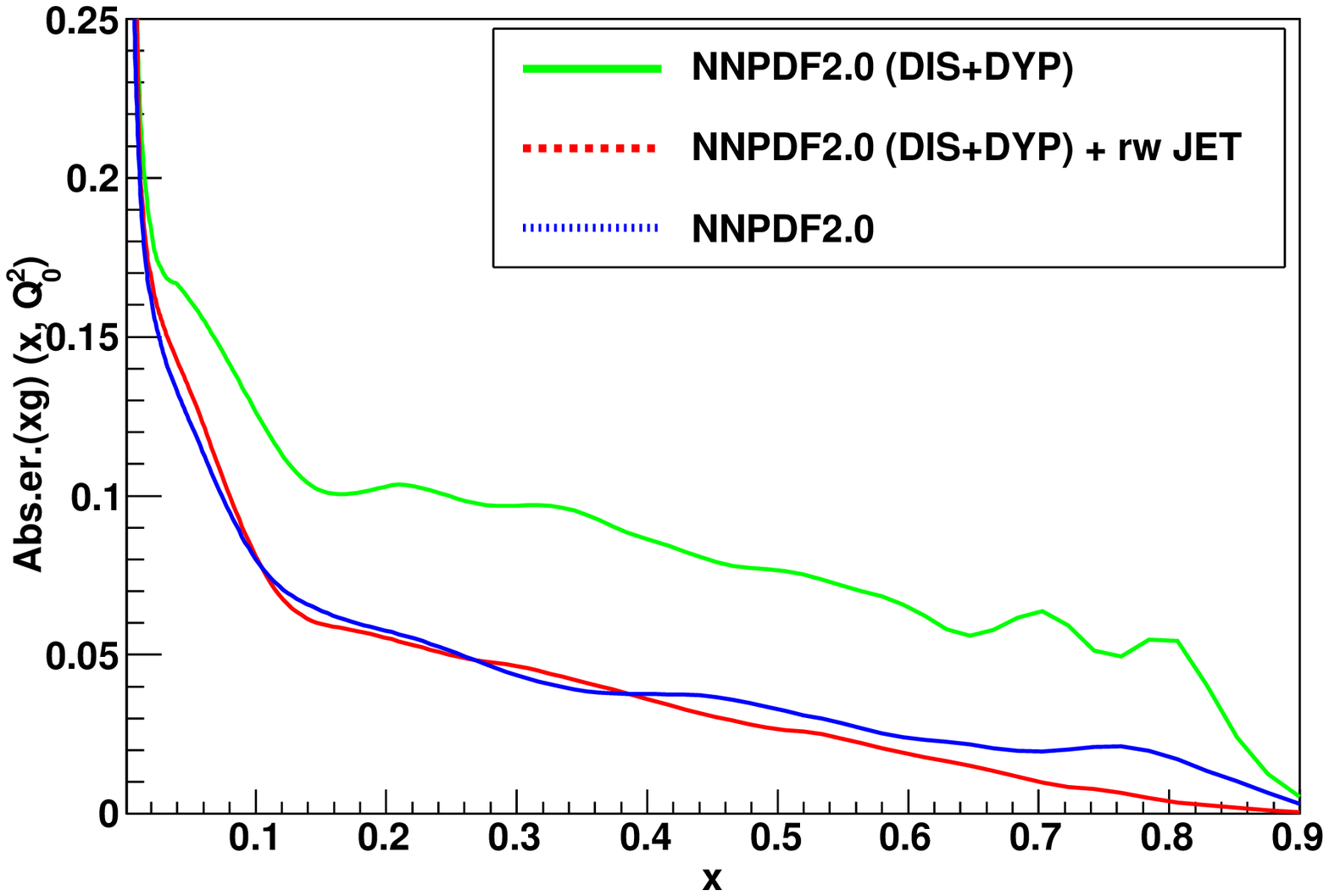}
    \end{center}
    \vskip-0.5cm
    \caption{\small The gluon distribution (left) and its uncertainty (right) of 
      the NNPDF2.0(DIS+DY) fit before and after reweighting with the inclusive jet data 
      compared to the refitted gluon from NNPDF2.0.}
    \label{fig:pdf-jets}
\end{figure}
\clearpage
\section*{Unweighting}

Once we have a reweighted PDF set, we would like to be able to produce a new PDF ensemble with the same probability distribution as a reweighted set, but without the need to include the weight information. A method of unweighting has therefore been developed, whereby the new set is constructed by deterministically sampling with replacement the weighted probability distribution. This means that replicas with a very small weight will no longer appear in the final unweighted set while replicas with large weight will occur repeatedly.

\begin{figure}[th!]
  \centering
  \epsfig{width=0.4\textwidth,figure=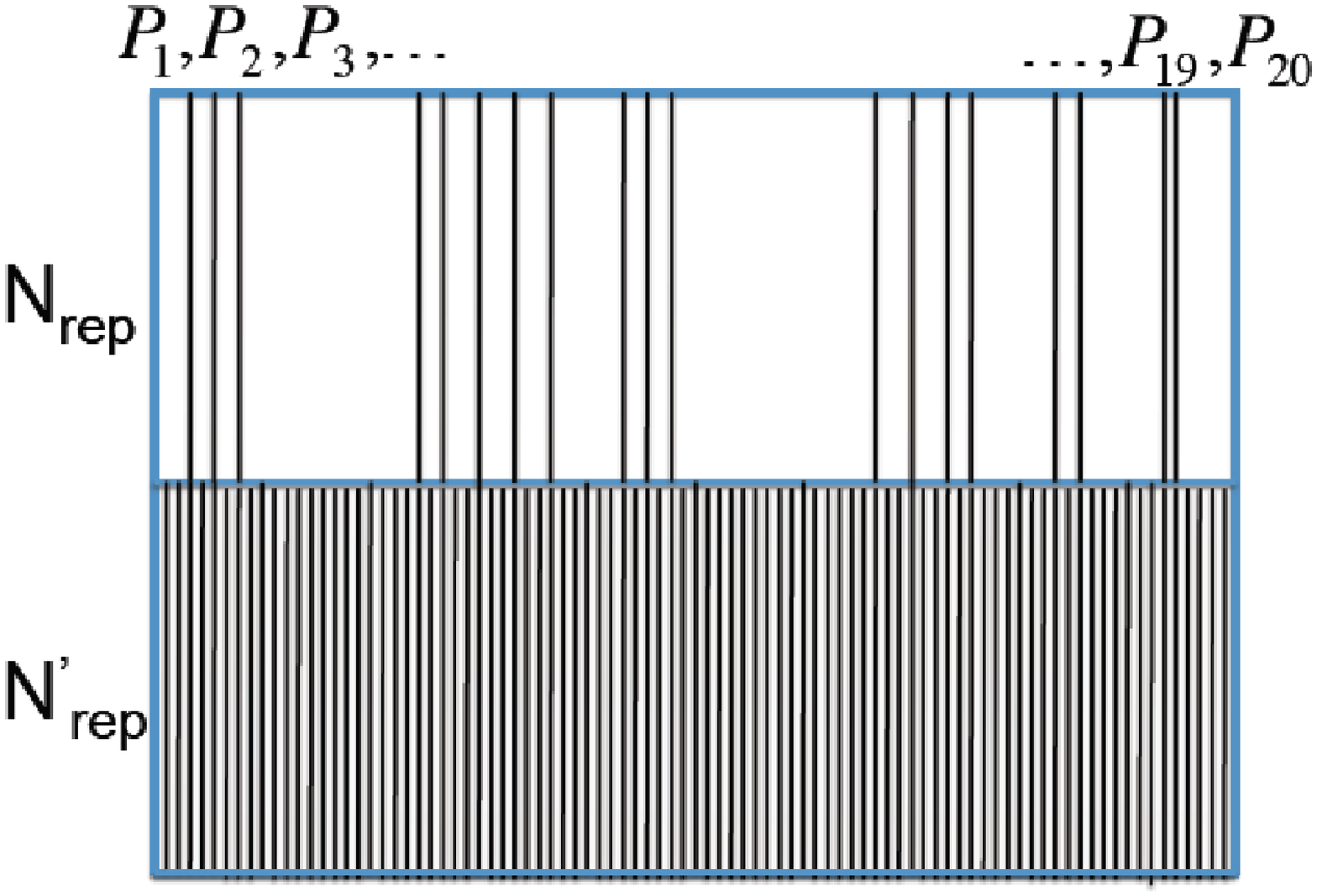,angle=-90}\\
  \vspace{-.5truein}
  \epsfig{width=0.4\textwidth,figure=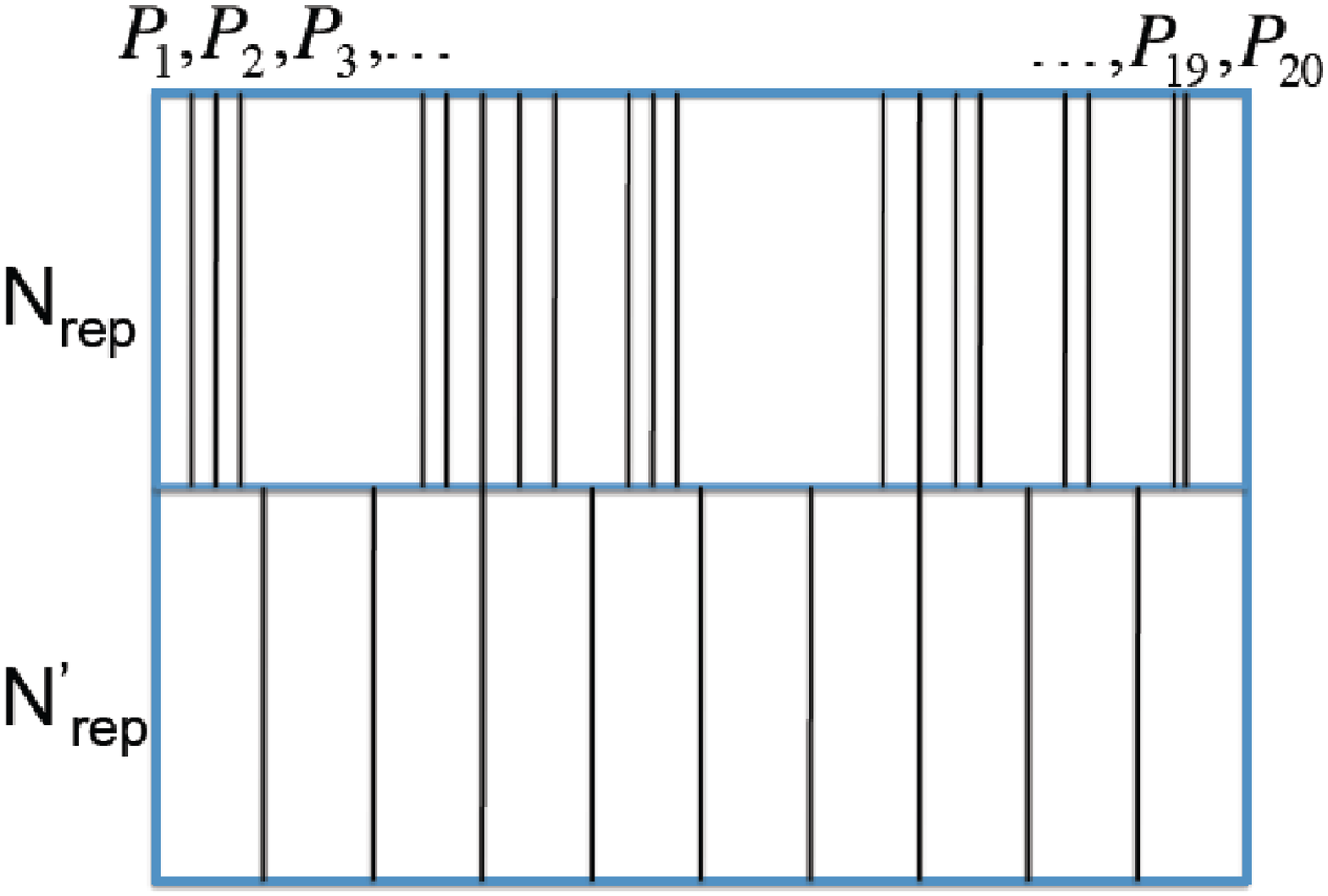,angle=-90}
  \caption{\small Graphical representation of the construction of a set
    of $N'_{rep}$ unweighted replicas from a set of $N_{rep}=20$
    weighted ones. Each segment is in one-to-one correspondence to a
    replica, and its length is proportional to the weight of the
    replica. The cases of  $N'_{rep}\gg N_{rep}$ (top) and
    $N'_{rep}=10$ (bottom) are shown.}
  \label{fig:unwplot}
\end{figure}
 
If we define the probability for each replica, and the probability cumulants as
$$
 p_k=\frac{w_k}{N_{rep}}\,\,\,\,\,\,\,\,\,\,\,\,\,\,\,\,\,\,\,\,\,\,P_k\equiv P_{k-1}+p_k=\sum_{j=0}^kp_j\, .
$$

we can quantitatively describe the unweighting procedure. Starting with $N_{rep}$ replicas with weights $w_k$, we determine $N_{rep}$ new weights $w'_k$:
$$
w'_k=\sum_{j=1}^{N'_{rep}}\theta\big(\frac{j}{N'_{rep}}-P_{k-1}\big)\theta\big(P_{k}-\frac{j}{N'_{rep}}\big)\, .
$$
These weights are therefore either zero or a positive integer. By construction they satisfy:
$$
N'_{\rm rep}\equiv \sum_{k=1}^{N_{\rm rep}} w'_k\, .
\label{eq:nprimedef}
$$
i.e the new unweighted set consists of $N'_{\rm rep}$ replicas, simply constructed by taking $w'_k$ copies of the $k$-th replica, for all
$k = 1, . . . ,N_{rep}\,$. This procedure is illustrated graphically in figure \ref{fig:unwplot}.

\section*{Testing Reweighting and Unweighting}
\begin{figure}[h]
  \centering
  \epsfig{width=0.99\textwidth,figure=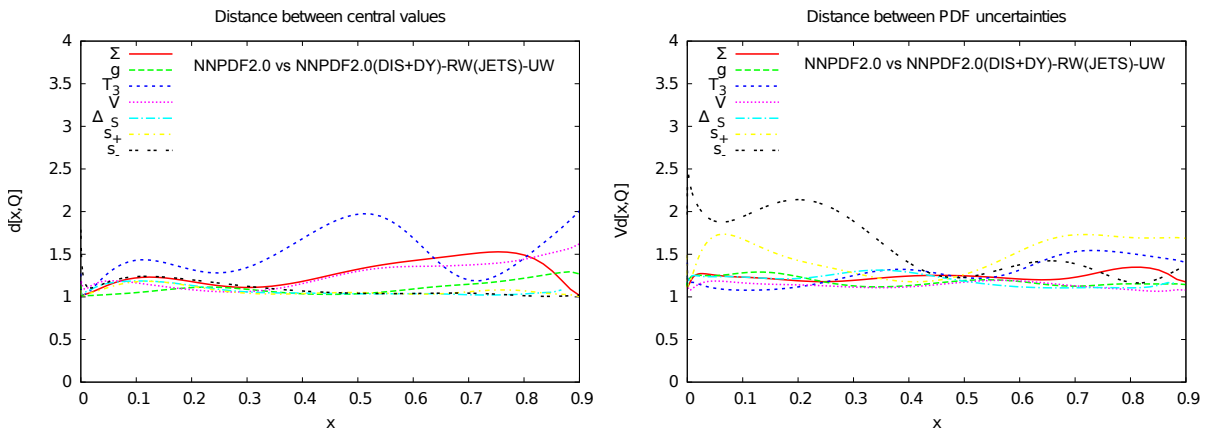}
  \caption{\small Distance between central values (left) and
    uncertainties (right) of the NNPDF2.0 PDFs and the NNPDF2.0 DIS+DY PDFs reweighted with Tevatron jet data and then unweighted. }
  \label{fig:distances-uw-jet}
\end{figure}

To verify the effectiveness of the reweighting procedure, we will show that including datasets by reweighting produces an ensemble of PDF replicas statistically equivalent to a full refit. We begin by producing a new NNPDF2.0 fit, including only DIS and Drell-Yan data. The data left out of the fit (Tevatron Run II inclusive jet data) is then reintroduced by reweighting. The resulting reweighted ensemble is then compared to the full NNPDF2.0 fit.
\\\\
In Figure \ref{fig:pdf-jets} we see the gluon PDF for the three sets; the prior fit NNPDF2.0(DIS+DY), the reweighted set NNPDF2.0(DIS+DY) with jet data included, and the refitted full set NNPDF2.0. The figure shows excellent agreement between the reweighted set and the full fit. Differences stay well below statistical fluctuations. 
\\\\
To obtain a more precise estimation of the statistical equivalence of the refitted and reweighted parton sets, and also to test the unweighting procedure, we may examine the statistical distances between the new unweighted distributions and the refitted set. The distance formulae are defined in Appendix A of Ref. \cite{Ball:2010de}. If two sets give a description of the same underlying probability distribution and so are statistically equivalent, the distance between them will fluctuate around a value of one. At $d\sim7$ the discrepancy between the two sets is at the one sigma level. In the case of the Tevatron jets reweighting exercise, we see in Figure \ref{fig:distances-uw-jet} that these distances oscillate around one. The reweighted set is therefore equivalent to the refit and there is no significant loss of accuracy in the unweighting procedure.

Having developed the unweighting procedure, we can perform another check on the consistency of the reweighting method. When adding more than one set of data by reweighting, our method must satisfy combination and commutation properties. Reweighting with both sets must be equivalent to reweighting with one, unweighting then reweighting with the other. Of course switching the order in which we reweight must produce an equivalent distribution.

 \begin{figure}[h]
   \centering
   \includegraphics[width=0.49\textwidth]{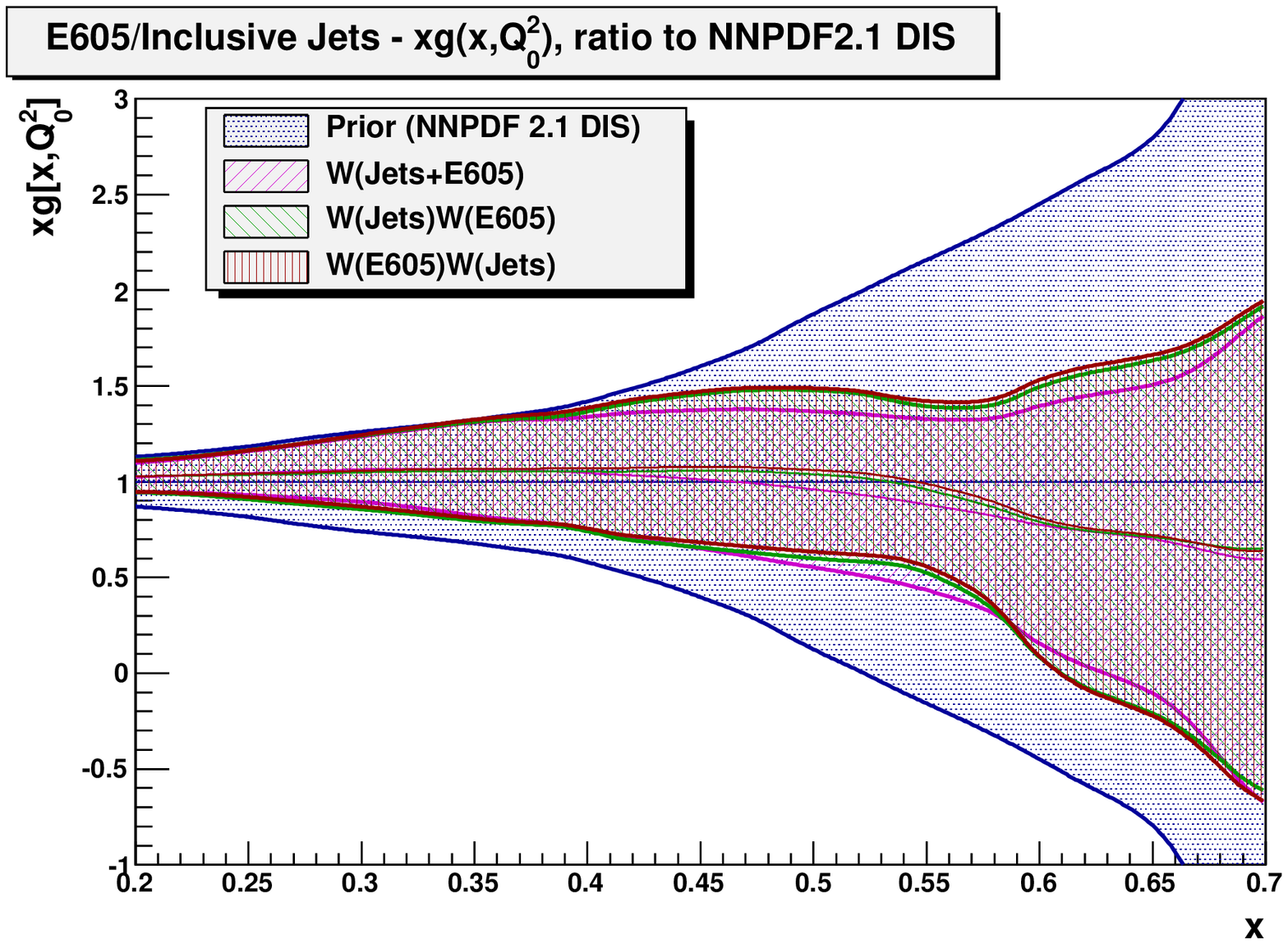}
   \includegraphics[width=0.49\textwidth]{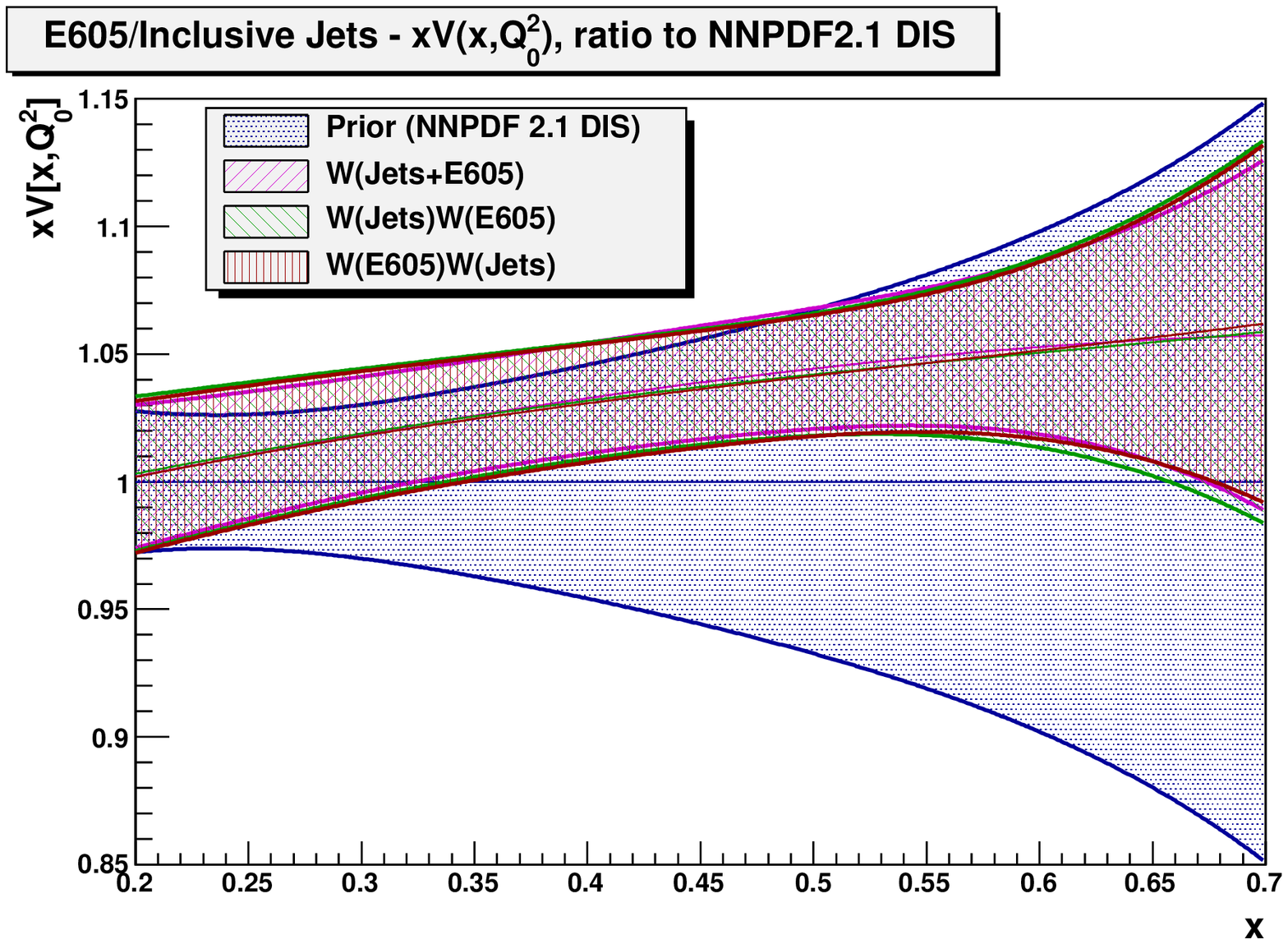}
   \caption{Multiple reweighting demonstration. Plots of gluon PDF(left) and valence PDF (right).}
   \label{fig:srw}
 \end{figure}

To check that our procedure satisfies these properties, we perform a test using the Tevatron jet data as the first dataset and E605 fixed target Drell-Yan data as the second. In Figure \ref{fig:srw} we compare the inclusion of the combined set with the inclusion of one set after the other. The result is clearly independent from the order in which the inclusion of single datasets is performed. A distance analysis performed on the three produced sets confirms that the reweighting method satisfies the combination and commutation requirements.

\begin{figure}[b!]
  \centering
  \epsfig{width=0.45\textwidth,figure=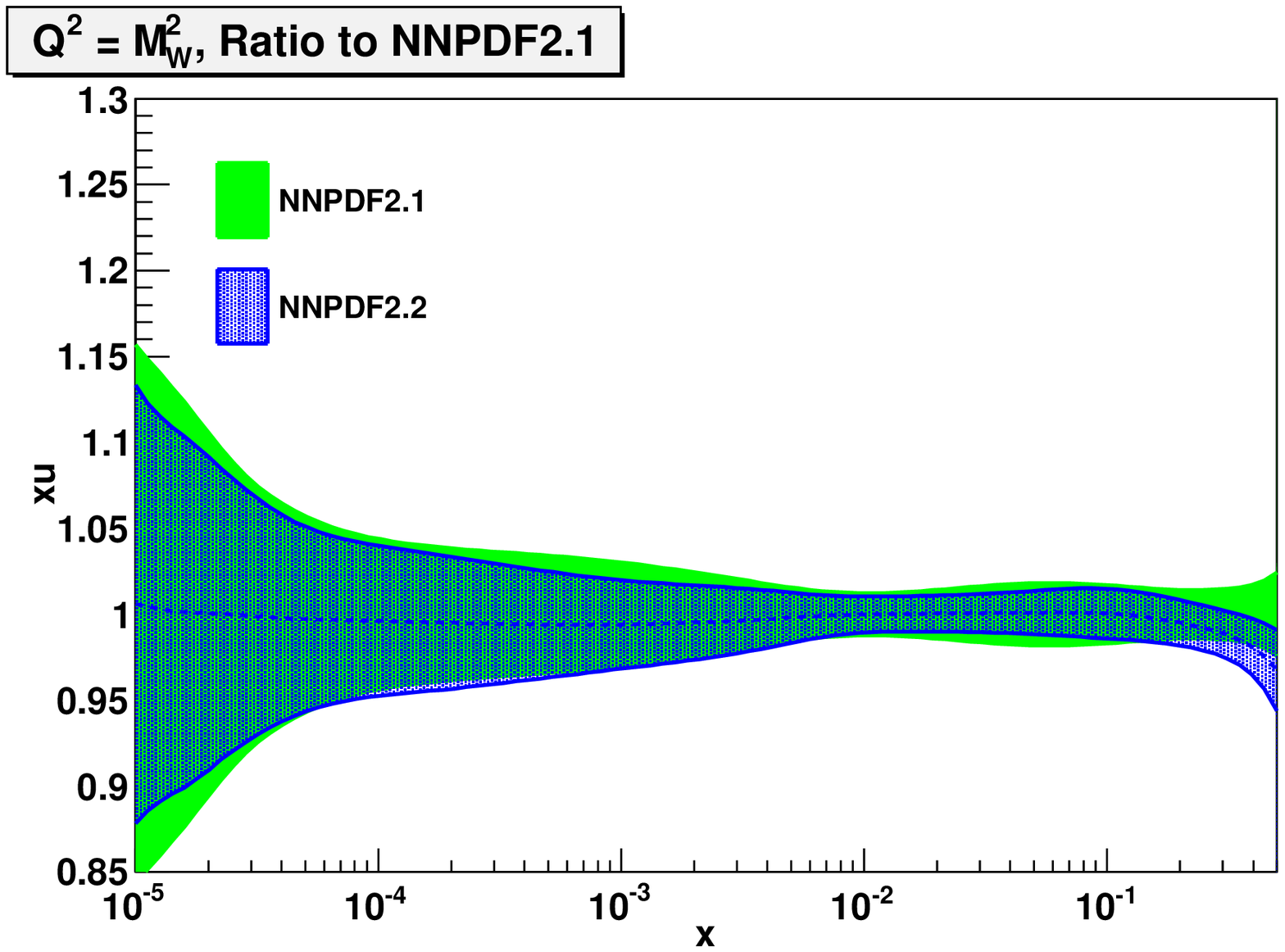}
  \epsfig{width=0.45\textwidth,figure=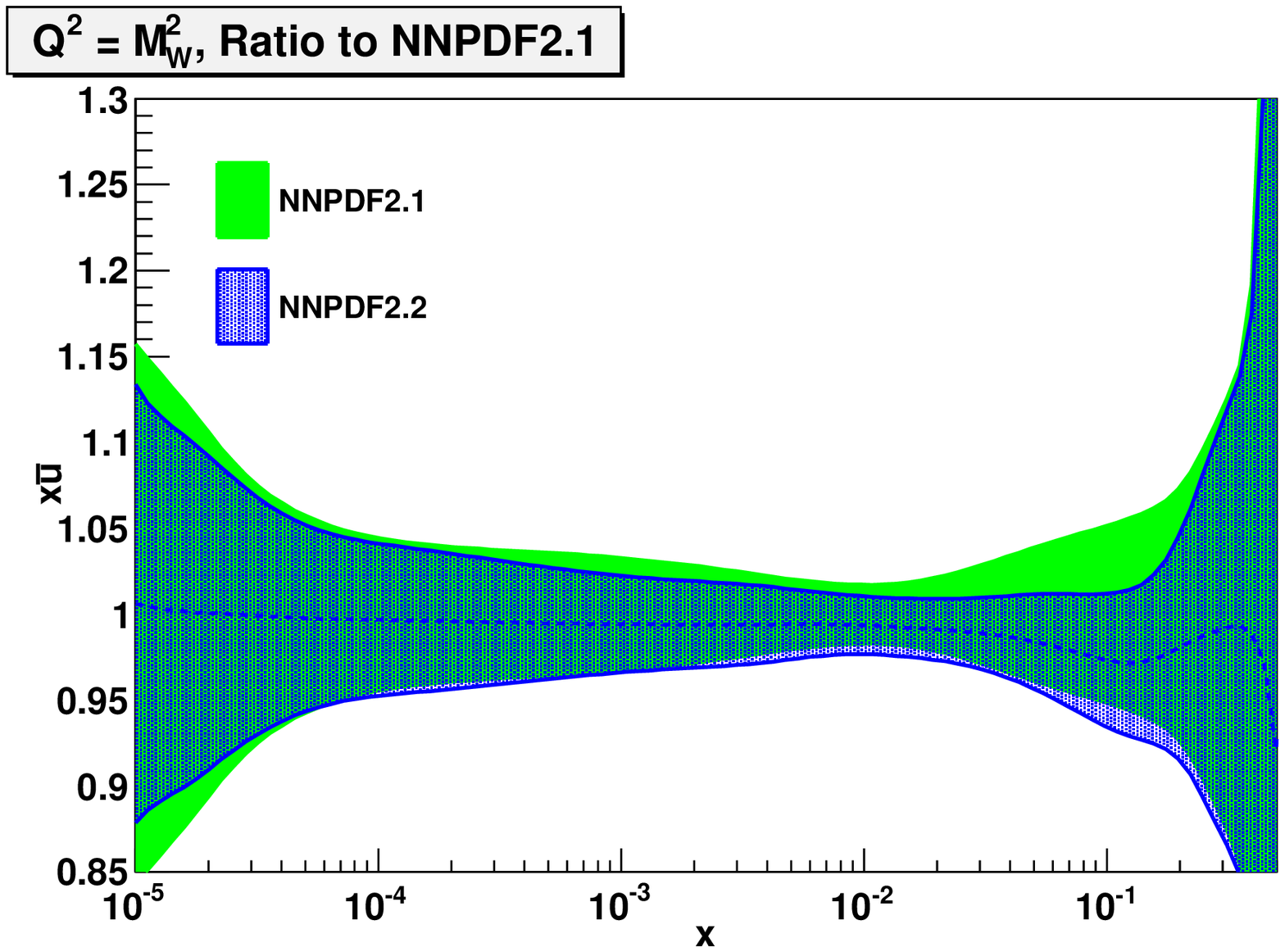}
  \epsfig{width=0.45\textwidth,figure=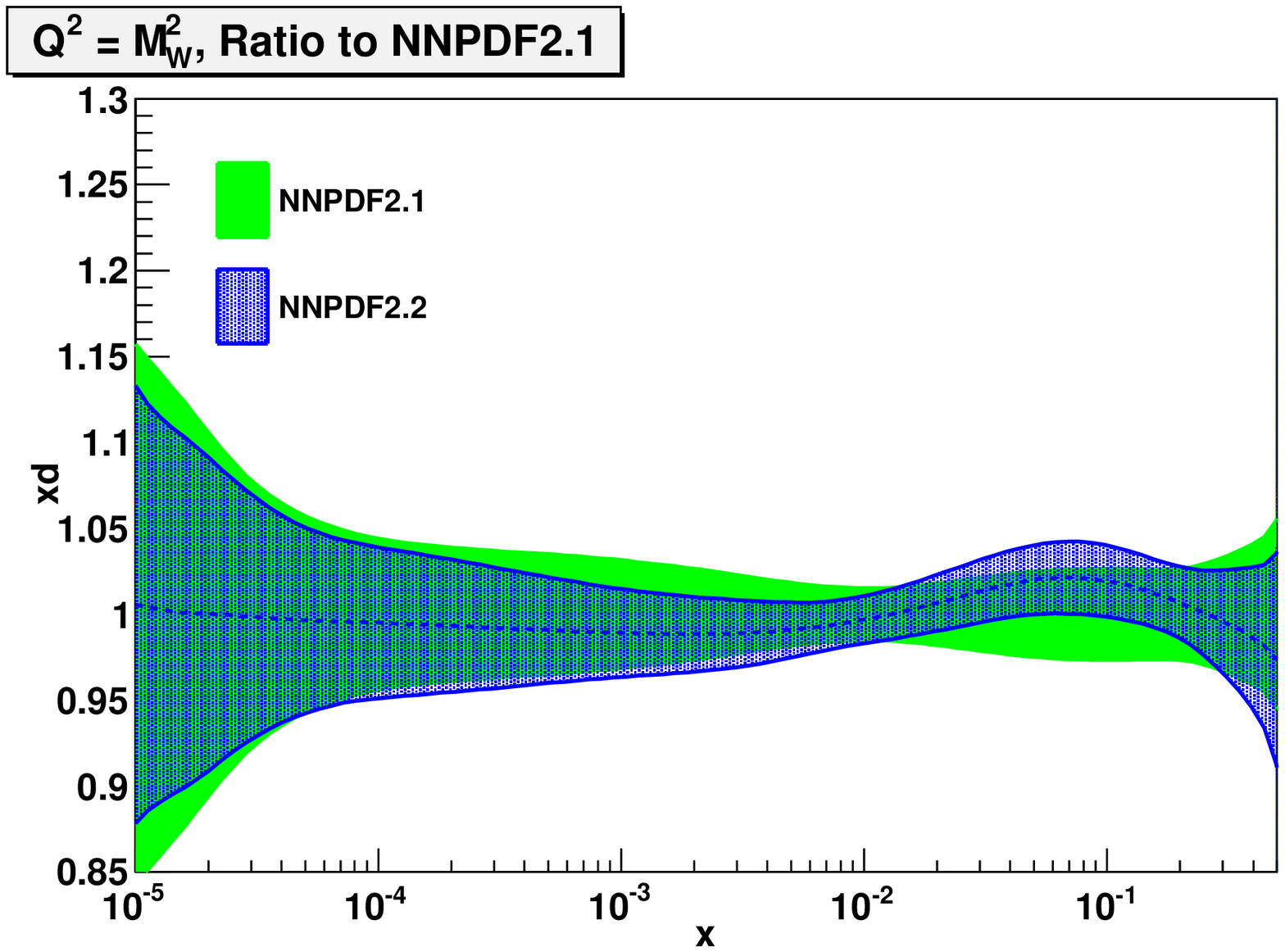}
  \epsfig{width=0.45\textwidth,figure=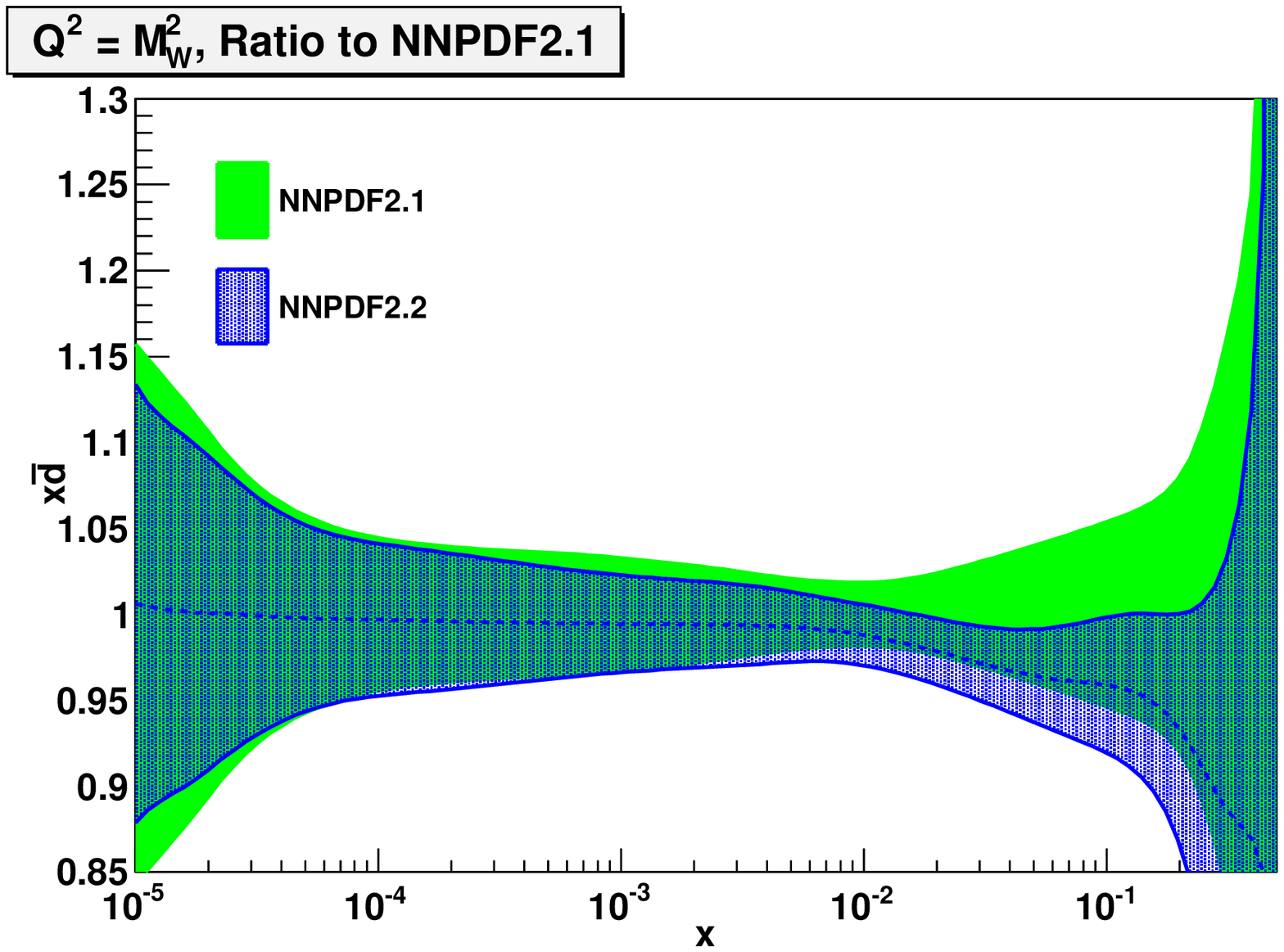}
  \caption{Comparison of light quark and antiquark distributions at the scale
    $Q^2=M_W^2$ from the global NNPDF2.1 and NNPDF2.2 global fits. 
    Parton densities are plotted normalized to the NNPDF2.1 central value.}
 \label{fig:pdf1-tev+lhc25}
\end{figure}

\newpage
\section*{NNPDF2.2}
The Bayesian reweighting method has been used to construct a new NNPDF parton set: NNPDF2.2\cite{rw2}. In this set we take as a prior ensemble the NNPDF2.1 fit and include by reweighting the W-lepton charge asymmetry measurements of the ATLAS, CMS and D0 collaborations.
\\\\
The NNPDF2.1 set provides a reasonable description of the new measurements, with $\chi^2_{tot}/N_{dat}=2.22$. After reweighting with the new data this improves to an excellent level of agreement with $\chi^2_{tot}/N_{dat}=0.81$. Having reweighted a prior set with $N_{\mathrm{rep}}=1000$ initial replicas, $181$ remain, indicating that the data provides a substantial constraint. Using the unweighting procedure outlined above, we have produced the new PDF set with $N_{\mathrm{rep}}=100$.
\\\\
Figure \ref{fig:pdf1-tev+lhc25} demonstrates the impact of the new data on the light quark and antiquark PDFs. The uncertainties are significantly constrained by the data in two main regions, there is a reduction of around $20\%$  at $x\sim10^{-3}$ and $30\%$ in the region $x\sim 10^{-2}$ to $x\sim 10^{-1}$. The overall fit quality improves slightly, from a total $\chi^2_{tot}/N_{dat}$ of 1.165 with NNPDF2.1 to $1.157$. The constraints demonstrated here are the first such constraints upon parton distributions from LHC data.
\\\\
Such constraints are particularly important given the discrepancies between global parton distribution fits in flavour separation at medium to large x. The W-lepton charge asymmetry data included here may prove useful in resolving some of these discrepancies. Future LHC data will no doubt provide further constraints upon PDFs in this region.

\end{document}